\documentclass[twocolumn,showpacs,amsmath,amssymb,superscriptaddress,pra]{revtex4}
\usepackage{graphicx}
\usepackage{dcolumn}
\usepackage{bm}
\def\btt#1{\texttt{\@backslashchar#1}}%
\DeclareRobustCommand\bblash{\btt{\@backslashchar}}%

\begin{document}

\title{Quantum Secure Communication via W States}

\author{Jaewoo Joo} \email{joogara@physics3.sogang.ac.kr}

\affiliation{Department of Physics, Sogang University, CPO Box 1142,
  Seoul 100-611, Korea}

\author{Jinhyoung Lee}

\affiliation{Institute of Quantum Information Processing and Systems
  University of Seoul, Seoul, Korea} 

\affiliation{School of Mathematics and Physics, The Queen's University,
  Belfast BT7 INN, United Kingdom}

\author{Jingak Jang}

\affiliation{National Security Research Institute, Taejon, Korea}

\author{Young-Jai Park}

\affiliation{Department of Physics and Basic Research Institute, Sogang
  University, Seoul 121-742, Korea}

\date{\today}

\begin{abstract}
  W states of multipartite systems are pair-wisely entangled, belonging
  to a different class from Greenberger, Horne, and Zeilinger states.
  Based on W states, we propose three variant protocols for quantum
  secure communication, {\em i.e.}, quantum key distribution, partial
  quantum secret sharing, and their synthesis. By the synthesis we mean
  that both quantum key distribution and partial quantum secret sharing
  are executed in a single protocol. For these protocols it is discussed
  how authorized communicators detect individual attacks by an
  eavesdropper.
\end{abstract}


\maketitle

\section{Introduction}

Quantum entanglement is at the heart of quantum information processes.
In particular bipartite entanglement has been extensively studied for
quantum key distribution \cite{Ekert91}, teleportation \cite{Bennett93},
entanglement swapping \cite{Zukowski93}, entanglement teleportation
\cite{Lee00}, conclusive teleportation \cite{MMM99}, and quantum
computation \cite{Neilsen00}. On the other hand multipartite
entanglement has been relatively less studied in terms of
information-theoretic aspects.

Recently the classification on the states of three qubits has been
proposed \cite{Dur99,Acin00,Acin01}. A typical class includes
Greenberger- Horne-Zeilinger (GHZ) states which exhibit nonlocality
among distant local observables \cite{GHZ90} and have nonvanishing
tangle \cite{Coffman00}. W states are pair-wisely entangled with no
tangle and they belong to the different class from GHZ states in the
sense that they can not be transformed to any GHZ states under the local
operation and classical communication (LOCC) \cite{Dur00}. These facts
suggest that W states have the physical properties considerably
different from GHZ states. W states have been rarely considered for
quantum information protocols, whereas GHZ states have been employed for
quantum cryptography among several distant parts
\cite{Hillery99,Kempe99,Scarani01}.

The nonlocality of W states were studied by explicitly considering
non-commuting observables \cite{Cabello01}.  The generation of W states
has been investigated with the physical models based on linear optical
elements \cite{Zeilinger97}, cavity QED \cite{Guo02}, and Heisenberg
$XY$ model \cite{Wang01}.

In this paper we propose three variant protocols by using W states of
three qubits which enable the quantum secure communication among three
distant persons. The protocols include quantum key distribution (QKD) in
the pair-wise way and partial quantum secret sharing (PQSS) among the
three persons. In addition we propose a synthesis protocol by which both
pair-wise QKD and PQSS are executed simultaneously when it is necessary.
It is discussed how authorized communicators detect individual attacks
by an eavesdropper.

\section{Characteristics of W states}

We consider briefly the known characteristics on W states of three
qubits and compare them with those of GHZ states. W states can not be
transformed to the GHZ state under LOCC, implying they belong to a
different class from the GHZ states \cite{Dur00}. 

A pure triseparable state of three qubits is defined by Schmidt-like
decomposition as \cite{Thapliyal99}
\begin{eqnarray}
  \label{eq:tss}
  |\phi_3\rangle = \sum_{i=1}^{2} \lambda_i |\alpha_i,\beta_i,\gamma_i\rangle
\end{eqnarray}
where $\{|\alpha_i\rangle\}$, $\{|\beta_i\rangle\}$ and
$\{|\gamma_i\rangle\}$ are orthonormal basis sets on the Hilbert spaces
for the three qubits, respectively, and $\lambda_i$ are positive.  The
triseparability is closely related to the tangle
$\tau=2\lambda_1\lambda_2$ \cite{Coffman00}. When
$\lambda_i=1/\sqrt{2}$, the triseparable state (\ref{eq:tss}) becomes a
maximal GHZ state with $\tau=1$. When the partial trace is performed
over one qubit, the reduced density operator of the other qubits can be
represented by a convex sum of product states. It implies that the
reduced density operator is separable.  On the other hand, it can be
shown by generalized Schmidt decomposition scheme \cite{Acin00} that W
states are not triseparable since they can not be represented in the
form of Eq.~(\ref{eq:tss}). Thus W states have no tangle, $\tau=0$. But
the reduced density operator for each pair among the three qubits is
inseparable because its partial transposition has at least one negative
eigenvalue. Thus, the W state is pair-wisely entangled with no tangle.

The nonlocality of W states can be revealed by three local observers.
Let $z^q$ and $x^q$ be the outcomes ($+$ or $-$) in
measuring $\hat{\sigma}_z$ and $\hat{\sigma}_x$ on qubit $q$
($q=A,B,C$). For the W state the local observables $z^q$ and
$x^q$ satisfy Einstein-Podolsky-Rosen (EPR) criterion of elements
of reality and should be predetermined before any measurement
\cite{Einstein35}.  However, such a predetermination is impossible
according to quantum mechanics. Recently, Cabello \cite{Cabello01} has
shown that quantum mechanics refutes EPR's elements of reality by the
proof similar to the GHZ's. It is notable that the refutation results
from the inferences by only two of three qubits while the inferences by
all three qubits are necessary for GHZ's.

Bell's theorem for W states of three qubits can be investigated also by
considering Clauser-Horne-Bell (CH-Bell) inequality \cite{Cabello01},
\begin{eqnarray}
\label{eq:ch}
&&-1 \le A_{11} - A_{12} - A_{21} - A_{22}\le 0, 
\end{eqnarray}
where $A_{11}=P(z^i = z_+, z^j = z_+)$ is the probability that two
qubits $i$ and $j$ among the three raise the outcomes of $z_+$ as they
are measured by $\hat{\sigma}_z$, and $A_{12}=P(z^i=z_+, x^j \ne x^k)$
is the probability that one qubit $i$ measured by $\hat{\sigma}_z$
raises the outcome of $z_+$ {\em and} the others $j$ and $k$ measured in
$\hat{\sigma}_x$ raise the outcomes opposite to each other. Similarly,
$A_{21}=P(x^i \ne x^k, z^j=z_+)$, and $A_{22}=P(x^i = x^j = x^k)$.

In particular consider a symmetric W state \cite{Coffman00},
\begin{eqnarray}
  \label{eq:ws}
  |W\rangle = \frac{1}{\sqrt{3}} \left( |z_-z_+z_+\rangle + |z_+z_-z_+\rangle +
   |z_+z_+z_-\rangle \right),
\end{eqnarray}
where $\{|z_+\rangle, |z_-\rangle\}$ is the set of the eigen states for
$\hat{\sigma}_z$ and the symbol ``$\otimes$'' for the direct product is
omitted unless any confusions arise.  For this symmetric W
state~(\ref{eq:ws}),
\begin{eqnarray}
\label{eq:probfw1}
P(z^i = z_+, z^j = z_+) &=& 1, \\
\label{eq:probfw2}
P(z^i=z_+, x^j \ne x^k) &=& 0, \\
\label{eq:probfw3}
P(x^i \ne x^k, z^j=z_+) &=& 0, \\
\label{eq:probfw4}
P(x^i = x^j = x^k) &=& 3/4.
\end{eqnarray}
The middle term in Eq.~(\ref{eq:ch}) is 1/4, implying the violation of
the CH-Bell inequality (\ref{eq:ch}).

\section{Quantum secure communication via W states}

\subsection{Pair-wise quantum key distribution}

Quantum key distribution (QKD) is a secure communication scheme by which
two distant persons have in common a secret key message via quantum
channels and classical communication. A QKD protocol includes three
basic steps: a) Encoding a key message on a quantum state, b)
transmitting the quantum system in the encoded state, and c) decoding
the key message from the state. Protocols for QKD may be divided into
two sets: protocols assisted by an entangled quantum channel and the
rest of protocols. The protocol suggested by Ekert (E91) \cite{Ekert91}
and that by Bennett and Brassard (BB84) \cite{Bennett84} are
representative of the two sets respectively.

In BB84, a sender encodes a key message on non-orthogonal states of a
quantum system ({\em e.g.}, photon polarization) which is directly
transferred to an authorized person, a receiver. The receiver can
retrieve the key message by a measurement on the state if the
measurement is approved through classical communication with the sender.
Security of BB84 has been investigated based on no cloning theorem
\cite{Bennett84,Shor00}.  Modified protocols have been proposed
\cite{Bennett92a}.

On the other hand, E91 utilizes an entangled EPR pair as a quantum
channel. Two distant persons share an EPR pair. Each person measures an
observable randomly chosen among three non-commuting observables. By
doing so, the two persons can have in common a key message if both
measurements are approved through classical communication. The
procedures for encoding and decoding are executed at the same time,
contrary to BB84. Security of E91 is based on Bell's theorem
\cite{Bell65}.  As the quantum channel is in an EPR state, the
authorized persons can see a violation of Bell's inequality from
measurement outcomes. Provided an eavesdropper enforces an
intercept-resend strategy to extract the key message \cite{Gisin02}, the
attempt breaks the entanglement of the quantum channel and thus the
Bell's inequality is not violated. The link between security of QKD and
Bell's inequality has been intensively discussed against individual
attacks by an eavesdropper \cite{Scarani01}.

Bennett {\em et al.}  \cite{Bennett92b} suggested a slightly variant
protocol (BBM92) from E91 with a simplified set of observables and
showed that BBM92 is actually equivalent to BB84 although they have
different characteristics on quantum channels. They investigated the
security by comparing {\em a priori} probabilities with {\em posteriori}
ones of outcomes instead of Bell's inequality.

We propose a pair-wise QKD protocol via W states of three qubits, which
is similar to BBM92 using EPR state of two qubits. Suppose that three
authorized persons, Alice, Bob, and Charlie, would like to perform a
secure communication in the pair-wise way such that every pair among the
three persons tries to have a key message and in particular the members
in the pair have in common the key message.

\begin{table}
\caption{Deciders and mutual relations among the outcomes in local
  measurements for pair-wise quantum key distribution}
\begin{ruledtabular}
\begin{tabular}{cccc}
Alice & Bob & Charlie  & decider \\ 
\colrule
$z_+$ & $x_+$ & $x_+$  & Alice  \\ 
$z_+$ & $x_-$ & $x_-$  & \\ 
$z_-$ & $x_+$ or $x_-$ & $x_+$ or $x_- $  & \\
\colrule
$x_+$ & $z_+$ & $x_+$  & Bob \\ 
$x_-$ & $z_+$ & $x_-$ &  \\ 
$x_+$ or $x_-$ & $z_-$ & $x_+$ or $x_-$ & \\ 
\colrule
$x_+$ & $x_+$ & $z_+$  & Charlie \\ 
$x_-$ & $x_-$ & $z_+$  & \\ 
$x_+$ or $x_-$ & $x_+$ or $x_-$ & $z_-$  &
\end{tabular}
\label{table1}
\end{ruledtabular}
\end{table}

Consider a composite system of three qubits which is in a symmetric W
state in Eq.~(\ref{eq:ws}). The three authorized persons share the three
qubits, one qubit for each person. As in E91 or BBM92 they acquire key
messages by performing local measurements. Each person randomly chooses
an observable out of $\hat{\sigma}_z$ and $\hat{\sigma}_x$. After his
measurement, he announces the axis but not the outcome. When Alice
measures along $d_1$ axis, Bob along $d_2$, and Charlie along $d_3$, the
set of the axes is denoted by $d_1$-$d_2$-$d_3$. We employ particular
sets of axes, {\em i.e.}, $x$-$x$-$z$, $x$-$z$-$x$, and $z$-$x$-$x$ for
the present pair-wise QKD. The probability to choose one of the sets is
3/8.

In Table~\ref{table1}, we present mutual relations among outcomes when
the authorized persons perform local measurements along a particular set
of axes. One who measured along $z$-axis decides whether it is possible
for the others to have in common a key message.  We call him a decider.
For example, consider the case of $x$-$x$-$z$. If Charlie obtains an
outcome $z_-$ in his local measurement along the $z$-axis, then Alice
and Bob have their pair in the product state $|z_+z_+\rangle$ and futher
they obtain outcomes randomly out of $x_+$ and $x_-$ in their local
measurements along the $x$-axis.  The outcomes are useless for the
present protocol and discarded as they exhibit no correlation.  However,
if Charlie obtains $z_+$, the pair that Alice and Bob share comes to be
in the maximally entangled state $(|z_-z_+\rangle +
|z_+z_-\rangle)/\sqrt{2}$. In the measurements along $x$-axis, Alice and
Bob obtain the same outcome of $x_+$ or $x_-$. They may now regard the
outcome as a key bit.  Similar procedures are applied to the rest of
$x$-$z$-$x$ and $z$-$x$-$x$.  The success probability in distributing a
key bit is 2/3 once a particular set of measurement axes is chosen among
$x$-$x$-$z$, $x$-$z$-$x$, and $z$-$x$-$x$.

The present pair-wise QKD protocol via W states is summarized as
following
\begin{enumerate}
\item[K.1~] Each of Alice, Bob, and Charlie chooses, at random, the axis
  of measuring instrument out of $x$- and $z$-axes.
\item[K.2~] Each person announces a bit information on the axis of his
  local measurement but not the outcome.
\item[K.3~] For the purpose of security one requests to announce their
  outcomes at random in the trials of distributing key messages.
\item[K.4a] All keep their outcomes if the set of the measurement axes
  is $x$-$x$-$z$, $x$-$z$-$x$, or $z$-$x$-$x$. Otherwise restart the
  protocol.
\item[K.4b] A decider who measured along $z$-axis tells the rest to
  regard their outcomes as a key bit if his outcome is $z_+$.  Otherwise
  restart the protocol.
\item[K.5~] Repeat the protocol until they have key bits as many as they want.
\item[K.6~] By obtaining frequencies of security-check events over the
  outcomes which were announced at the step K.3, verify the security of
  quantum channel against attacks by an eavesdropper (See
  Sec.~\ref{sec:security} for details). If the errors are larger than
  permitted, throw away the key bits which have been obtained so far.
\end{enumerate}

Overall success probability $P_s$ in obtaining a key bit is given by
multiplying together the probability of choosing a particular set of
measurement axes and the success probability of distributing a key bit
for a given particular set of axes.  The number of distributed key bits
$K_t$ is given as
\begin{eqnarray}
\label{eq:ntkb}
K_t = P_s N_{e}
\end{eqnarray}
where $P_s$ is an overall success probability and $N_e$ an effective
number of trials. In the pair-wise QKD $P_s=1/4$ and $N_e=N-M$ where $N$
is a number of total trials in the protocol and $M$ a number of trials
for the purpose of security. For a given $K_t$, the protocol requires
$n_q$ qubits such that
\begin{eqnarray}
  \label{eq:noq}
  n_q=\frac{3K_t}{P_s(1+ M/N)}.
\end{eqnarray}
In the limit of $N \rightarrow \infty$, $M/N \rightarrow 0$ and $n_q
\rightarrow 3K_t/P_s$. Thus the protocol requires $12$ qubits per a key
bit at average. On the other hand, the protocol of E91 has the overall
success probability $P_s=2/9$ and it requires $9$ qubits per a key bit.

\subsection{Partial quantum secret sharing}

Quantum secret sharing is a key distribution protocol among $N$ persons
in such a way that one's key message can be retrieved by the others only
if they cooperate all together. Variant protocols have been proposed by
using multipartite GHZ states \cite{Hillery99,Scarani01} and bipartite
EPR states \cite{Karlsson99,Tittel01}.

Here we consider a partial quantum secret sharing protocol (PQSS) using
W states of three qubits. The procedure for PQSS among three persons is
similar to the pair-wise QKD. The difference is the set of measurement
axes, {\em i.e.}, $z$-$z$-$z$ is employed.

Suppose Alice, Bob, and Charlie would like to perform QSS by sharing
three qubits in a symmetric W state (\ref{eq:ws}). Each person randomly
chooses an observable out of $\hat{\sigma}_z$ and $\hat{\sigma}_x$ as
done in the pair-wise QKD. After his measurement, he announces the axis
but not the outcome. In the case of $z$-$z$-$z$, he may regard the
outcome as a key bit. More explicitly, consider a case that Bob and
Charlie are expected to retrieve Alice's key message in their
cooperation. If Alice has the outcome $z_+$, then Bob and Charlie have
opposite outcomes out of $z_+$ and $z_-$. Otherwise Both have the same
outcome $z_+$. When Bob and Charlie cooperate so as to collect their
outcomes, they can correctly deduct Alice's key bit.  We note that, if
Bob or Charlie obtains $z_-$, he will realize that the other and Alice
have the same outcome $z_+$. In the case he can deduct Alice's key bit
without help of the other.  It will be done in the probability of 1/3.
The fact implies that Bob and Charlie may have partial information on
Alice's key bit.  However, Bob and Charlie must still cooperate to
retrieve completely Alice's key bits. In the sense we call the present
protocol a PQSS.

Although the actual key sharing is performed in the case of $z$-$z$-$z$,
the authorized persons must choose $\hat{\sigma}_x$ for their
observables as well for the purpose of security. If they chose only
$z$-axis for PQSS, an eavesdropper could extract all information by
measuring along $z$-axis.

The protocol of PQSS has steps in common with the pair-wise QKD
protocol: Most steps in the pair-wise QKD protocol are applicable to
PQSS but the step K.4 is modified as
\begin{enumerate}
\item[S.4] All keep their outcomes if the set of the measurement axes is $z$-$z$-$z$.
  Otherwise restart the protocol.
\end{enumerate}

In PQSS we have the overall success probability $P_s=1/8$ which is
determined by the probability of choosing $z$-$z$-$z$. Due to the
arguments below Eq.~(\ref{eq:ntkb}), $24$ qubits are necessary to share
a key bit at average in the limit of $N \rightarrow \infty$.  On the
other hand, HBB99 with $P_s=1/2$ needs $6$ qubits per a key bit.

\subsection{Synthesis of QKD and PQSS}

The proposed protocols for QKD and PQSS via W states overlap the
following procedures, a) performing local measurements for an observable
out of $\hat{\sigma}_z$ and $\hat{\sigma}_x$ at random and b) checking security by
obtaining frequencies in security-check events. They differ in choosing
a set of measurement axes and their sets of axes are mutually exclusive.
This observation enables a synthesis protocol by which both QKD and PQSS
are performed simultaneously. That is, if authorized persons choose a
set of axes out of $x$-$x$-$z$, $x$-$z$-$x$, and $z$-$x$-$x$, their
outcomes are used for QKD and, if they choose $z$-$z$-$z$, their
outcomes are used for PQSS.  We construct the synthesis protocol by
inserting a new step B.3$'$ between K.3 and K.4 in the protocol of
pair-wise QKD as
\begin{enumerate}
\item[B.3$'$] If the set of the measurement axes is $z$-$z$-$z$, go to
  the step S.4 and otherwise go to K.4.
\end{enumerate}

In the synthesis protocol, the probability to perform QKD is not equal
to that for PQSS. Of successful trials for QKD or PQSS, 2/3 and 1/3 are
involved in QKD and PQSS respectively.  The overall success probability
for the synthesis protocol is obtained by summing those for QKD and PQSS
so that $P_s=3/8$. Due to the arguments below Eq.~(\ref{eq:ntkb}), in
the limit of $N\rightarrow \infty$, the necessary number of qubits is
$n^{syn}_q=8$ per a key bit at average.
In order to compare the synthesis protocol, let us consider a separate
protocol performed by E91 for QKD and HBB99 for QSS in probabilities 2/3
and 1/3 respectively. The necessary number of qubits is $n^{sep}_q = 8$,
which is equal to $n^{syn}_q$.

\section{Security in the proposed protocols}
\label{sec:security}

How is it guaranteed that the distributed key messages are secure
against attacks by an eavesdropper?  An eavesdropper, Eve, may have
several kinds of strategies to attack \cite{Gisin02}. A systematic
method for security is not known against all possible strategies. We
shall consider simple individual attacks such that Eve performs a
unitary operation on a composite system of her auxiliary qubit and one
of three qubits, which are involved in a secure communciation, and she
tries to extract some information by measuring her auxiliary qubit
\cite{Bennett92b,Scarani01,Gisin02}. We show that authorized persons can
detect these attacks as they examine frequencies of their outcomes.

Suppose Eve tries to extract some information from a qubit for Charlie.
Preparing her qubit in the state $|z_+\rangle$ for a simplicity and
intercepting the qubit on the way for Charlie, Eve applies to the
composite system of her and Charlie's qubit a unitary operation of
\begin{eqnarray}
  \hat{U}_{CE} |z_+z_+\rangle_{CE} &=& |z_+z_+\rangle_{CE} \nonumber \\
  \hat{U}_{CE} |z_-z_+\rangle_{CE} &=& \cos \phi |z_-z_+\rangle_{CE} + \sin \phi
  |z_+z_-\rangle_{CE}
\end{eqnarray}
where $\phi \in [0,\pi/2]$ characterizes the strength of Eve's attack.
This unitary operation was considered to study the optimal individual
attack by Eve \cite{Niu99}.  Due to eavesdropping, the four-qubit state
of $A$, $B$, $C$, and $E$ reads
\begin{eqnarray}
  |W'\rangle &=&\frac{1}{\sqrt{3}} \big(|z_-z_+z_+z_+\rangle +
    |z_+z_-z_+z_+\rangle \nonumber \\
    & & + \cos \phi |z_+z_+z_-z_+\rangle + \sin \phi |z_+z_+z_+z_-\rangle \big).
\end{eqnarray}
This state is no longer symmetric. The probabilities given in
Eqs.~(\ref{eq:probfw1}) to (\ref{eq:probfw4}) come to vary with respect
to the observing persons. In particular $P(z^C=z_+, x^A
\ne x^B) \ne P(z^A=z_+, x^B \ne x^C)$ in
Eq.~(\ref{eq:probfw2}).

It is clear that the trial of Eve's eavesdropping changes not only the
state of quantum channel but also its probabilities in
Eqs.~(\ref{eq:probfw1}) to (\ref{eq:probfw4}). For example, consider the
event that one person has the outcome $z_i=z_+$ and the others have the
opposite outcomes $x_j \ne x_k$ once the authorized persons choose one
of $x$-$x$-$z$, $x$-$z$-$x$, and $z$-$x$-$x$ for their measurement axes.
The probability of such an event to occur vanishes as given in
Eq.~(\ref{eq:probfw2}) if there is no eavesdropping. However, as Eve
tries the individual attack, the probabilty of such an event becomes
$(1-\cos^2\phi)/6$ if Charlie chooses $z$-axis and $(1-\cos\phi)/3$ if
he does $x$-axis. This implies that such an event is expected to occur
in the probability
\begin{eqnarray}
  \label{eq:probws2m}
  \bar{P} = \frac{(1-\cos\phi)(5+\cos\phi)}{18}
\end{eqnarray}
which is obtained by averaging the probabilities over Charlie's two
cases.  Keeping it in mind that such an event never happens if there is
no attack by Eve, the authorized persons can detect Eve's eavesdropping
by checking whether such an event occurs.  We call, by a security-check
event, an event which can be used to detect Eve's eavesdropping. For the
security of QKD, PQSS, and the synthesis, we employ the event that one
person has the outcome $z_i=z_+$ and the others have the opposite
outcomes $x_j \ne x_k$. This security event has an advantage over others
such that they never occur if no attack by Eve and their occurrence
indicates directly Eve's attack.

Subensemble of local-measurement outcomes suffices to examine the
security of the proposed protocols.  One of three persons may request at
random to announce their outcomes by $M$ times among $N$ trials for
distributing key messages. We call $M$ a number of security trials.
Based on $M$ sets of outcomes, they obtain the frequency of the
security-check event and verify whether there is an attack by Eve. This
procedure is appended to the protocols as steps K.3 and K.6.

\section{Remarks}

We have proposed three variant protocols of quantum key distribution,
partial quantum secret sharing, and their synthesis based on W states,
which have the different nonlocal characteristics from GHZ states.  We
have shown that these protocols are secure against the simple individual
attacks by an eavesdropper.

\acknowledgments This work has been supported by the BK21 Project No.
D-1099 of the Korea Ministry of Education.  J. L. has been supported in
part by the Korean Ministry of Science and Technology through the
Creative Research Initiatives Program under Contract No. 00-C-CF01-C-35
and by the Post-doctoral Fellowship Program of Korea Science \&
Engineering Foundation (KOSEF)

\end{document}